\documentstyle[12pt]{article}

\def\hybrid{\topmargin -20pt	\oddsidemargin 0pt
	\headheight 0pt	\headsep 0pt
        \textwidth 6.35in
        \textheight 9.65in
	\marginparwidth .875in
	\parskip 5pt plus 1pt	\jot = 1.5ex}
\hybrid

\def\baselinestretch{1.2}

\catcode`\@=11

\def\marginnote#1{}
\newcount\hour
\newcount\minute
\newtoks\amorpm
\hour=\time\divide\hour by60
\minute=\time{\multiply\hour by60 \global\advance\minute by-\hour}
\edef\standardtime{{\ifnum\hour<12 \global\amorpm={am}%
	\else\global\amorpm={pm}\advance\hour by-12 \fi
	\ifnum\hour=0 \hour=12 \fi
	\number\hour:\ifnum\minute<10 0\fi\number\minute\the\amorpm}}
\edef\militarytime{\number\hour:\ifnum\minute<10 0\fi\number\minute}
\def\draftlabel#1{{\@bsphack\if@filesw {\let\thepage\relax
   \xdef\@gtempa{\write\@auxout{\string
      \newlabel{#1}{{\@currentlabel}{\thepage}}}}}\@gtempa
   \if@nobreak \ifvmode\nobreak\fi\fi\fi\@esphack}
	\gdef\@eqnlabel{#1}}
\def\@eqnlabel{}
\def\@vacuum{}
\def\draftmarginnote#1{\marginpar{\raggedright\scriptsize\tt#1}}

\def\draft{\oddsidemargin -.2truein
	\def\@oddfoot{\sl preliminary draft \hfil
	\rm\thepage\hfil\sl\today\quad\militarytime}
	\let\@evenfoot\@oddfoot	\overfullrule 3pt
	\let\label=\draftlabel
	\let\marginnote=\draftmarginnote
   \def\@eqnnum{(\theequation)\rlap{\kern\marginparsep\tt\@eqnlabel}%
\global\let\@eqnlabel\@vacuum}  }

\def\preprint{\twocolumn\sloppy\flushbottom\parindent 2em
	\leftmargini 2em\leftmarginv .5em\leftmarginvi .5em
	\oddsidemargin -.5in	\evensidemargin -.5in
	\columnsep .4in	\footheight 0pt
	\textwidth 10.in	\topmargin  -.4in
	\headheight 12pt \topskip .4in
	\textheight 6.9in \footskip 0pt
	\def\@oddhead{\thepage\hfil\addtocounter{page}{1}\thepage}
	\let\@evenhead\@oddhead	\def\@oddfoot{}	\def\@evenfoot{} }

\def\numberbysection{\@addtoreset{equation}{section}
	\def\theequation{\thesection.\arabic{equation}}}

\def\underline#1{\relax\ifmmode\@@underline#1\else
	$\@@underline{\hbox{#1}}$\relax\fi}

\def\titlepage{\@restonecolfalse\if@twocolumn\@restonecoltrue
\onecolumn
     \else \newpage \fi \thispagestyle{empty}\c@page\z@
	\def\thefootnote{\fnsymbol{footnote}} }

\def\endtitlepage{\if@restonecol\twocolumn \else \newpage \fi
	\def\thefootnote{\arabic{footnote}}
	\setcounter{footnote}{0}}  

\catcode`@=12
\relax

\def\np{Nucl. Phys. \/}
\def\pl{Phys. Lett. \/}
\def\figcap{\section*{Figure Captions\markboth
	{FIGURECAPTIONS}{FIGURECAPTIONS}}\list
	{Figure \arabic{enumi}:\hfill}{\settowidth\labelwidth{Figure
999:}
	\leftmargin\labelwidth
	\advance\leftmargin\labelsep\usecounter{enumi}}}
 \relax
\def\tablecap{\section*{Table Captions\markboth
	{TABLECAPTIONS}{TABLECAPTIONS}}\list
	{Table \arabic{enumi}:\hfill}{\settowidth\labelwidth{Table
999:}
	\leftmargin\labelwidth
	\advance\leftmargin\labelsep\usecounter{enumi}}}
 \relax
\def\reflist{\section*{References\markboth
	{REFLIST}{REFLIST}}\list
	{[\arabic{enumi}]\hfill}{\settowidth\labelwidth{[999]}
	\leftmargin\labelwidth
	\advance\leftmargin\labelsep\usecounter{enumi}}}
 \relax
\makeatletter
\newcounter{pubctr}
\def\publist{\@ifnextchar[{\@publist}{\@@publist}}
\def\@publist[#1]{\list
	{[\arabic{pubctr}]\hfill}{\settowidth\labelwidth{[999]}
	\leftmargin\labelwidth
	\advance\leftmargin\labelsep
	\@nmbrlisttrue\def\@listctr{pubctr}
	\setcounter{pubctr}{#1}\addtocounter{pubctr}{-1}}}
\def\@@publist{\list
	{[\arabic{pubctr}]\hfill}{\settowidth\labelwidth{[999]}
	\leftmargin\labelwidth
	\advance\leftmargin\labelsep
	\@nmbrlisttrue\def\@listctr{pubctr}}}
 \relax
\makeatother
\newskip\humongous \humongous=0pt plus 1000pt minus 1000pt

\newif\ifdtup

\relax
\hybrid

\def\thefootnote{\fnsymbol{footnote}}
\def\be{\begin{equation}}
\def\ee{\end{equation}}
\def\ba{\begin{eqnarray}}
\def\ea{\end{eqnarray}}

\def\cm{{\cal M}}
\def\cg{{\cal G}}
\def\ov{\overline}
\def\gev{{\rm \; GeV}}
\def\tev{{\rm \; TeV}}
\def\str{\rm \; Str \;}
\def\mpl{M_{\rm P}}
\def\mun{M_{\rm U}}
\def\msu{M_{\rm SUSY}}
\def\simlt{\stackrel{<}{{}_\sim}}
\def\str{{\rm Str \,}}
\def\zzbar{(z, \overline{z})}

\def\np{Nucl. Phys. \/}
\def\pl{Phys. Lett. \/}

\begin{document}
\renewcommand{\theequation}{\thesection.\arabic{equation}}
\newcommand{\beq}{\begin{equation}}
\newcommand{\eeq}[1]{\label{#1}\end{equation}}
\newcommand{\ber}{\begin{eqnarray}}
\newcommand{\eer}[1]{\label{#1}\end{eqnarray}}
\begin{titlepage}
\begin{center}

\hfill CERN-TH/95-293\\
\hfill LPTENS-95/48\\
\hfill hep-th/9512034\\

\vskip .15in

{\large \bf Superstring Extension of the Standard Model and
Gravitation}
\vskip .3in

{\bf  Costas Kounnas\footnote{On leave from Ecole
Normale Sup\'erieure, 24 rue Lhomond, F-75231, Paris, Cedex 05,
FRANCE.}}\\
\vskip
 .2in

{\em Theory Division, CERN,\\ CH-1211,
Geneva 23, SWITZERLAND} \footnote{e-mail addresse:
KOUNNAS@NXTH04.CERN.CH}\\

\vskip .2in

\end{center}

\vskip .15in

\begin{center} {\bf ABSTRACT } \end{center}
\begin{quotation}\noindent

Some of the four-dimensional Superstring solutions provide a
consistent framework for a Supersymmetric Unification of all
interactions including gravity.  A class of them extends successfully
 the validity of the standard model up to the string scale ${\cal
O}(10^{17})~GeV$. We stress the importance of string corrections
which are relevant for low energy ${\cal O}(1)~TeV$ predictions of
gauge and Yukawa couplings as well as the spectrum of the
supersymmetric particles after supersymmetry breaking.

\end{quotation}

\vspace*{.5cm}
\begin{center}
{\it Talk given at the }\\
\vskip 0.2cm
{\it SUSY-95 International Workshop}\\
{\it ``Supersymmetry and Unification of Fundamental Interactions"} \\
{\it Palaiseau, 15-19 May 1995} \\
{\it and at the} \\
{\it  Four Seas Conference }\\
{\it Trieste, June 25-July 1, 1995.}
\end{center}
\vspace*{1cm}

\vskip 1.0cm
\vskip 0.5cm
\begin{flushleft}
CERN-TH/95-293 \\
November 1995\\
\end{flushleft}

\end{titlepage}
\vfill
\eject
\def\baselinestretch{1.2}
\baselineskip 16 pt
\noindent
\section{Introduction and Motivations}

If one tries to extend the validity of an effective field theory to
energy scales much higher than its characteristic mass scale, and
quantum corrections appear carrying positive powers of the cut-off
scale $\Lambda$, one is faced with a scale hierarchy problem. The
typical example is the scale hierarchy problem [\ref{hierarchy}] of
the Standard Model (SM) of strong and electroweak interactions, seen
as a low-energy effective field theory. When the SM is extrapolated
to cut-off scales $\Lambda \gg 1 \tev$, there is no symmetry
protecting the mass of the elementary Higgs field, and therefore the
masses of the weak gauge bosons, from large quantum corrections
proportional to $\Lambda$. The most popular solution to the scale
hierarchy problem of the SM is [\ref{les}] to extend the latter to a
model with global $N=1$ supersymmetry, effectively broken at a scale
$\msu \simlt 1 \tev$. These extensions of the SM, for instance
[\ref{mssm}] the Minimal Supersymmetric Standard Model (MSSM), can be
safely extrapolated up to cut-off scales much higher than the
electroweak scale, such as the supersymmetric unification scale $\mun
\sim
10^{16} \gev$, the string scale $M_S \sim 10^{17} \gev$, or the
Planck scale $\mpl \equiv G_N^{-1/2} / \sqrt{8 \pi} \simeq 2.4 \times
10^{18} \gev$.

To see in a more quantitative way the properties of supersymmetry
that guarantee the stability of the hierarchy of scales against
quantum corrections in the MSSM and its variants, it is useful  take
a closer look  to the one-loop effective potential of a generic
theory using  a momentum cut-off $\Lambda$, [\ref{cww}]
$$
V_1 = V_0 + {1 \over 64 \pi^2} \str \cm^0 \cdot \Lambda^4 \log
{\Lambda^2 \over \mu^2} + {1 \over 32 \pi^2} \str \cm^2 \cdot
\Lambda^2 +
$$
\be
\label{veff}
{1\over 64 \pi^2} \str \cm^4 \log {\cm^2 \over \Lambda^2}  + \ldots
\,
,
\ee
where the dots stand for $\Lambda$- independent contributions, $\mu$
is the scale parameter, and
\be
\label{straccia}
\str \cm^n \equiv \sum_i (-1)^{2J_i} (2 J_i + 1) m_i^n
\ee
is a sum over the $n$-th power of the various field- dependent mass
eigenvalues $m_i$, with weights accounting for the number of degrees
of freedom and the statistics of particles of different spin $J_i$.
In eq.~(\ref{veff}),
$V_0$ is the classical potential, which in the case of the SM (and of
the MSSM) should contain mass terms at most of the order of the
electroweak scale.

The quantum correction to the vacuum energy with the highest degree
of ultraviolet divergence is the $\Lambda^4$ term, whose coefficient
$\str \cm^0$ is always field- independent, and equal to the number of
bosonic minus fermionic degrees of freedom. Being field- independent,
this term can affect
the discussion of the cosmological constant problem (when the theory
is coupled to gravity), but does not affect the discussion of the
hierarchy problem. Anyway, this term is always absent in
supersymmetric theories, which possess equal numbers of bosonic and
fermionic degrees of freedom.

The second most divergent term in eq.~(\ref{veff}) is the
quadratically divergent contribution which is proportional to $\str
\cm^2$. In the SM, $\str \cm^2$ depends on the Higgs field, and
induces a quadratically divergent contribution to the Higgs squared
mass, the well-known source of the  hierarchy
problem. An early attempt to get rid of the quadratically divergent
one-loop contributions to the SM Higgs squared mass consisted
[\ref{veltman}] in imposing the mass relation $[(\partial^2 /
\partial \varphi^2) \str \cm^2 (\varphi) ]_{\varphi = v} = 0$;
neglecting the light fermion masses, this amounts to requiring $3
m_H^2 +
6 m_W^2 + 3 m_Z^2 - 12 m_t^2 = 0$. It is clear (for recent
discussions, see e.g. [\ref{einhorn}] and references therein) that
such a requirement is modified at
higher orders in perturbation theory, since it amounts to a
relation among the dimensionless couplings of the SM that is not
stable under the renormalization group. A more satisfactory solution
of the problem is provided by $N=1$ global supersymmetry. For
unbroken $N=1$ global supersymmetry, $\str \cm^n$ is identically
vanishing for any $n$, due to the fermion-boson degeneracy within
supersymmetric multiplets. The vanishing of
$\str \cm^2$ persists, as a field identity, if global supersymmetry
is spontaneously broken in the absence of anomalous $U(1)$ factors
[\ref{fgp}]. Indeed, to keep the scale hierarchy stable it is
sufficient that supersymmetry
breaking does not reintroduce field- dependent quadratically
divergent contributions to the vacuum energy. This still allows for a
harmless, field- independent quadratically divergent contribution to
the effective potential, and is actually used to classify the
so-called soft supersymmetry- breaking terms~[\ref{gg}]. In the case
of softly broken supersymmetry, the $\Lambda^2$ term of
eq.~(\ref{veff}) only contributes to the cosmological constant. With
a typical mass splitting $\msu$ within the MSSM supermultiplets, the
logarithmic term in
eq.~(\ref{veff}) induces corrections to the Higgs mass terms (before
minimization), which are at most $O(\msu^2)$: the hierarchy is then
stable if $\msu \simlt {\cal O}(1) \tev$.

Without any further assumption, the predictability of the
supersymmetric standard model is rather weak, since one must include
all soft- breaking terms that are consistent with the
SU(3)$\times$SU(2)$\times$U(1)
gauge symmetry.  This means that the mass spectra of the squarks and
sleptons are unknown parameters, and that the only restriction on
them is just to be smaller than
or equal to $\msu \leq {\cal O}(1)~{\rm TeV}$.

In order to go further and give a more restricted mass spectrum for
the
squarks, sleptons, etc., more assumptions must be introduced in the
would-be
fundamental theory.  One interesting possibility is the unification
of the strong and electroweak gauge couplings, at high energy scales
$(E \geq 10^{16}~{\rm GeV})$. It is very interesting that the SUSY
Grand Unification idea works perfectly [\ref{unif}] and predicts
correctly the present experimental value for $sin^2\theta_W(m_Z) =
0.2334 \pm 0.0008$, once we set $\msu$ around the TeV scale and use
the actual values for $\alpha_s(m_Z) = 0.118 \pm 0.008$ and
$\alpha^{-1}_{em}(m_Z) = 127.9 \pm 0.2$.  The bottom to tau mass
ratio $(m_b/m_{\tau} \simeq 2.8)$ is also predicted correctly;  the
proton lifetime is predicted to be above the present experimental
limits $(\tau^{SUSY}_p \simeq 6 \times 10^{34}~{\rm years},~
\tau^{exp}_p  \geq 6.8 \times 10^{32}~{\rm years})$, due to the
relatively large grand-unification scale $M_{U} \simeq 1.1 \times
10^{16}$~GeV.

Within the SUSY grand-unification framework, the number of
independent
supersymmetry soft-breaking mass parameters is drastically reduced at
the unification
scale to the well known by now supersymmetry-breaking parameters
parameters.

\centerline{$m_0$, $m_{1/2}$, $A$, $B$ and $\mu$, at $M_U$,}

Due to the large radiative corrections going from the $M_U$ to the
$m_Z$ scale, the simple tree-level relations get modified
[\ref{susyrad},\ref{susyradk}]. The gauge radiative corrections
give positive contribution to the (mass)$^2$ parameters of the
squarks sleptons and Higgses. On the other hand, the large Yukawa
couplings of the third generation, give large negative  radiative
corrections to the (mass)$^2$ of the scalars and especially to the
one of the two Higgses. When the third generation
Yukawa couplings are sufficiently large, $m^2_H$ can become negative
at $m_Z$
scale; the SU(2)$\times$U(1) gauge symmetry is then spontaneously
broken by the gauge and Yukawa radiative corrections
[\ref{susyrad},\ref{susyradk}].
The precise value of the $h_t$ and $h_b$ couplings, which are
necessary to break the SU(2)$\times$U(1) at the correct scale, depend
on
the soft supersymmetry- breaking parameters. Obviously, the radiative
breaking mechanism requests a special relation between the gauge
couplings and the Yukawa couplings, and thus it demands the top quark
to be heavy enough to make the $m^2_{H}$ mass parameter negative.  A
detailed analysis, made twelve years ago [\ref{susyrad},
\ref{susyradk}] and more recently [\ref{susyradf}] shows that the
radiative breaking holds in certain regions of the soft breaking
parameters space provided that the top Yukawa coupling is
sufficiently large. Furthermore the radiative breaking mechanism
implies definite relations among gauge and Yukawa couplings. It is
impossible to understand such  relations
in the framework of non-supersymmetric or $N=1$ supersymmetric field
theory.
The only thing we may say at this point is that  they are necessary
to explain the SU(2)$\times$ U(1) breaking at the experimentally
known scale.  In more
fundamental theories like superstring theories, there are super-
unification relations among gauge and Yukawa couplings which are
compatible with the radiative breaking mechanism (see later) and the
present experimental data.

\section{Soft Breaking Terms from $N$=1 Supergravity}

In order to go beyond the {\em technical solution} of the hierarchy
problem in renormalizable softly broken supersymmetry, we must move
in to a more fundamental theory where the soft breaking
terms are generated via a spontaneous breaking of supersymmetry.
{\em The only possible candidate} for such a theory is $N=1$
supergravity coupled to
gauge and matter fields [\ref{cfgvp}]. In contrast with the
case of global supersymmetry the spontaneous breaking of local
supersymmetry is not incompatible with vanishing vacuum energy. In
$N=1$ supergravity, the spin $2$ graviton has for superpartner the
spin $3/2$ gravitino, and the only consistent way of breaking
supersymmetry is {\em spontaneously}, via the super-Higgs
mechanism. One is then bound to interpret the MSSM as an effective
low-energy
theory derived from a spontaneously broken supergravity.
The scale of soft supersymmetry breaking in the MSSM, $\msu$, is
related (in a
model-dependent way) to the gravitino mass $m_{3/2}$, which sets the
scale of the spontaneous breaking of local supersymmetry. One might
naively think that, whatever mechanism breaks local supersymmetry and
generates the hierarchy $m_{3/2} \ll \mpl$, the condition $\msu \sim
m_{3/2} \simlt 1 \tev \ll \mpl$ remains sufficient to guarantee the
stability of such a hierarchy against quantum corrections. To explain
why this expectation is generically incorrect [\ref{fkpz}], we need
first to review some general facts about spontaneously broken $N=1$
supergravity.

Even barring higher-derivative terms, the general structure of $N=1$
supergravity still allows for a large amount of arbitrariness
[\ref{cfgvp}]. First, one is free to choose the field content.
Besides the gravitational
supermultiplet, containing as physical degrees of freedom the
graviton and the gravitino, one has a number of vector
supermultiplets, whose physical degrees of freedom are the spin $1$
gauge bosons $A_\mu^a$ and the spin $1/2$ Majorana gauginos
$\lambda^a$, transforming in the adjoint representation of the chosen
gauge group. One is also free to choose the number of chiral
supermultiplets, whose physical degrees of freedom are spin $1/2$
Weyl fermions $\chi^I$ and complex spin $0$ scalars $z^I$, and their
transformation properties under the gauge group. Furthermore, one has
the freedom to choose a real gauge-invariant K\"ahler function
\be
\label{gfun}
\cg (z,\ov{z}) = K (z,\ov{z}) + \log |w(z)|^2 \, ,
\ee
where $K$ is the K\"ahler potential whose second derivatives
determine the kinetic terms for the fields in the chiral
supermultiplets, and $w$ is the (analytic) superpotential. One can
also choose a second (analytic) function $f_{ab} (z)$, transforming
as a symmetric product of adjoint representations of the gauge group,
which determines the kinetic terms for the fields in the vector
supermultiplets, and in particular the gauge coupling constants and
axionic couplings,
\be
g_{ab}^{-2} =  {\rm Re \,} f_{ab}  \, ,
\;\;\;\;\;
\theta_{ab} =  {\rm Im \,} f_{ab}  \, .
\ee
Once the functions $\cg$ and $f$ are given, the full supergravity
Lagrangian is specified. In particular (using here and in the
following the standard supergravity mass units in which $\mpl=1$),
the classical scalar potential reads
$$
V = V_F + V_D
$$
$$
V_F = e^\cg \left( \cg^I \cg_I - 3 \right)
$$
\be
\label{pot}
V_D= {[({\rm Re \,}
f)^{-1}]^{ab} \over 2} \left( \cg_I T^{I}_{a \;\;\; \ov{J}}
\ov{z}^{\ov{J}} \right) \left( z^L T^{\;\;\;\;\; \ov{K}}_{b \, L}
\cg_{\ov{K}} \right) \, .
\ee

In our notation, repeated indices are summed, unless otherwise
stated;
we use Hermitian generators, $[(T_a)^I_{\;\;\; \ov{J}}]^\dagger$ =
$T^{J}_{a \;\;\;\;\; \ov{I}}$; derivatives of the K\"ahler function
are denoted by $\partial \cg / \partial z^I \equiv \partial_I \cg
\equiv \cg_I$ and $\partial \cg / \partial \ov{z}^{\ov{I}} \equiv
\partial_{\ov{I}} \cg \equiv \cg_{\ov{I}}$; and the K\"ahler metric
is $\cg_{I \ov{J}}$ = $\cg_{\ov{J} I}$ = $K_{I \ov{J}}$ = $K_{\ov{J}
I}$.
The inverse K\"ahler metric $\cg^{I \ov{J}}$, such that $\cg^{I
\ov{J}} \cg_{\ov{J} K}$ = $\delta^I_K$, can be used to define
\be
\cg^I \equiv \cg^{I \ov{J}} \cg_{\ov{J}}
\, ,
\;\;\;\;\;
\cg^{\ov{I}} \equiv \cg_{J}\cg^{J \ov{I}}
\, .
\ee
Notice that the $D$-term part of the scalar potential is always
positive semi-definite, $V_D \ge 0$, as in global supersymmetry.
However, in contrast with global supersymmetry, the $F$-term part of
the scalar potential is not semi- positive definite in general. On
the
one hand, this allows for spontaneous supersymmetry breaking with
vanishing classical vacuum energy, as required by consistency with a
flat background. On the other hand, the requirement of vanishing
vacuum energy imposes a non-trivial constraint on the structure of
the theory,
\be
\label{vacuum}
\langle \cg^I  \cg_I \rangle = 3
\;\;\;\;\;
{\rm if}
\;\;\;\;\;
\langle V_D \rangle = 0  \, .
\ee

The order parameter of local supersymmetry breaking in flat space
is the gravitino mass,
\be
\label{mgrav}
m_{3/2}^2 \zzbar = |w(z)|^2 e^{K \zzbar}  \, ,
\ee
which depends on the vacuum expectation values of the scalar fields
of the theory, determined in turn by the condition of minimum vacuum
energy. In the following, we shall assume that the $D$ breaking
is absent at tree level, as is the case in all interesting
situations.
For convenience, we shall also classify the fields as $z^I \equiv (
z^\alpha,z^i)$, where the fields with Greek indices have
non-vanishing auxiliary fields  $
\cg^\alpha \ne 0$ and thus participate to the supersymmetry breaking;
the fields
with small Latin indices have vanishing  $ \cg^i  = 0$. With these
conventions,
eq.~(\ref{vacuum}) can be split as
\be
\langle \cg^\alpha  \cg_\alpha \rangle = 3 \, ,
\;\;\;\;\;
\langle \cg^i  \cg_i \rangle = 0 \, ,
\ee
The above $F$-  breaking defines the goldstino direction in terms of
the chiral fermions $\chi^I$,
\be
\label{gold}
\tilde{\eta} =  e^{\cg \over 2} \cg_{\alpha} \chi^{\alpha} \, .
\ee
and the $N=1$ local supersymmetry is spontaneously broken on a flat
background.  The coefficient of the one-loop quadratically divergent
contributions to the vacuum energy, associated to this breaking, is
given by [\ref{grk}]
\be
\label{genstr}
\str \cm^2 \zzbar = 2 \, Q \zzbar \, m_{3/2}^2 \zzbar
\, ,
\ee
where
\be
\label{qexpr}
Q \zzbar = N_{TOT} - 1 - \cg^I  \zzbar H_{I \ov{J}} \zzbar
\cg^{\ov{J}} \zzbar \, ,
\ee
$$
H_{I \ov{J}} \zzbar = R_{I \ov{J}} \zzbar + F_{I \ov{J}} \zzbar
$$
$$
R_{I \ov{J}} \zzbar \equiv \partial_{I} \partial_{\ov{J}}
\log \det \cg_{M \ov{N}} \zzbar \, ,
$$
\be
\label{ficci}
F_{I \ov{J}} \zzbar  \equiv \partial_{I} \partial_{\ov{J}}
\log \det {\rm Re \,} [ f_{ab} (z) ] \, .
\ee
Clearly, the only non-vanishing contributions to $\str \cm^2$ come
from the field directions $z^\alpha$ for which $\langle \cg^\alpha
\cg_\alpha \rangle \ne 0$ (not summed). In eq.~(\ref{ficci}), $R_{I
\ov{J}}$ is the Ricci tensor of the K\"ahler manifold for the chiral
multiplets, whose total number is denoted by $N_{TOT}$; $F_{I
\ov{J}}$ has also a geometrical interpretation, since the way it is
constructed from the gauge field metric is very similar to the way
$R_{I \ov{J}}$ is constructed from
the K\"ahler metric. It is important to observe that both $R_{I
\ov{J}}$ and $F_{I \ov{J}} $ {\em do not depend at all} on the
superpotential of the theory, but only depend on the
metrics for the chiral and gauge superfields. This very fact allows
for the
possibility that, for special geometrical properties of these two
metrics, the dimensionless quantity $Q \zzbar$ may turn out to be
field-independent and hopefully vanishing.

In a general spontaneously broken $N=1$ supergravity, the
non-vanishing of $Q \zzbar$ induces, at the one-loop level, a
contribution to the vacuum energy quadratic in the cut-off $\Lambda$.
This leads to a very
uncomfortable situation, not only in relation with the cosmological
constant problem (a vacuum energy of order $m_{3/2}^2 \Lambda^2$
cannot be cancelled by any physics at lower energy scales)
but also in relation with the gauge hierarchy problem, which asks for
a gravitino mass not much larger than the electroweak scale. Since
$m_{3/2} \zzbar$ is a field- dependent object, and its expectation
value must arise from minimizing the vacuum energy, quadratically
divergent loop corrections to the latter may generically destabilize
[\ref{fkpz}, \ref{destab}] the desired hierarchy $m_{3/2} \ll
\Lambda$,
attracting the gravitino mass either to $m_{3/2}=0$ (unbroken
supersymmetry) or to $m_{3/2} \sim \Lambda$ (no hierarchy).
This destabilization problem cannot be solved just by moving from the
cut-off regulated supergravity to the quantum supergravity defined by
four- dimensional superstrings [\ref{fds1}--\ref{fds6}], since the
only practical difference will be to replace the cut-off scale
$\Lambda$ by an effective scale of order $M_S$. In a generic
supergravity theory, we still have the freedom to evade this problem,
by postulating the existence of an extra sector of the theory, which
gives an opposite contribution to $Q$, so that $Q + \Delta Q = 0$.
Such a request,
however, is very unnatural, and implies a severe fine- tuning among
the parameters of the old theory and of the extra sector. In string-
derived supergravities, the possibility of such a cheap way out is
lost, since all the degrees of freedom of the theory are known and
the total contribution to $Q$ is well defined. We no longer have the
freedom to compensate a non-zero $Q$ by modifying the theory!

{}From the previous discussion, it is clear that a satisfactory
solution of the hierarchy problem ($m_{3/2} \ll \mpl$), and the
perturbative stability of the flat background, at least up to
${\cal O} (m_{3/2}^4)$ corrections, require the vanishing of $Q
\zzbar$ [\ref{fkpz}]. It is also clear that, if such a solution
exists, this will put strong constraints on the scalar and gauge
metrics, see eqs.~(\ref{genstr})--(\ref{ficci}). Even if one arrange
such relations by hand they will not be preserved in the quantum
(gravitational) level. In my opinion {\em there is no any hope} to
find such relations in the framework of supergravity field theories.
In  order to have any hope to find any solution to the above problem,
it is necessary to have a  framework in which  the quantum
corrections including the gravitational ones are well defined. The
only candidate theory we have so far, that contains a consistent
theory of quantum gravity is {\em superstring} theory.

Before moving to the superstring effective supergavities, I would
like
to present at this point a special class of $N=1$ supergravity
models, the so called {\em no-scale models} [\ref{nsm}, \ref{fkpz}].
The main property of no-scale models is that the classical vacuum
energy is identically equal to zero for all scalar field directions
$z^a$ which participate  to the breaking of supersymmetry with $\cg^a
\ne 0$.
\be
V(z^a; z^i)\geq 0,~~{\rm and}~~V_{min}(z^a; z^i)=0~~{\rm
for~any}~z^a.
\ee
The vacuum is classically degenerate  even after the supersymmetry
breaking.
In these models the  supersymmetry breaking scale $m_{3/2} \zzbar$ is
undetermined at the classical level.  At the quantum
level, however, the $z^a$-effective potential is deformed, and in
general one finds preferred values for the vev's of $z^a$ and thus
for $m_{3/2}\zzbar$.  There are two generic possibilities after the
quantum corrections; either,
$$
i)~m_{3/2}\zzbar \simeq O(\mpl)~,
$$
or
\be
ii)~m_{3/2}\zzbar \simeq O(Q_0)~,
\ee
where  $Q_0$ is a
dimensional transmutation scale, at which the mass$^2$ of the
SM--Higgs become negative, $m_H^2(Q_0) \sim 0$ and which is created
by the gauge and
Yukawa radiative  corrections.  The first possibility may take place
if the
quantum gravity corrections are large and do not respect the flatness
properties of the $z^a$-effective potential.  This possibility cannot
be examined in any theory in which the quantum gravity is not
consistently quantized. Only in the string framework can one check
this possibility.

For the same reasons the second possibility, $m_{3/2}\zzbar$ $
\simeq$ ${\cal O}(Q_0)$ can exist only under the assumption that the
quantum gravitational effects do not change drastically  the
$z^a$-flatness properties of the effective potential.  This
requirement strongly restricts the
fundamental theory at energy scales of order $E\simeq \mpl$.
Even in the string framework, there is no definite answer yet, even
though there are some indications that this is the case in a certain
class of
string- induced no-scale models.  {\em Assuming} this property of the
quantum gravitational corrections and that $Q\zzbar=)$, we can
examine in detail the effects of the gauge and Yukawa interactions at
all energy scales $E < \mpl$.  It turns out that $m_{3/2}\zzbar$ is
always attracted to an infrared critical scale near $Q_0$, which is
hierarchically smaller than $\mpl$; $m_{3/2}\zzbar = {\cal
O}(10^{-15})$$\mpl$ [\ref{nsm}, \ref{fkpz}].
Here I will just sketch this mechanism, since details of the
no-scale radiative breaking mechanism can be found in the literature
[\ref{nsm}, \ref{fkpz}].
 It is important that $Q_0$ exists for any value of $m_{3/2}\zzbar
\le Q_0$ and is almost $z^a$-independent.  For all $z^a$ such that
$m_{3/2}\zzbar \gg Q_0$,
$m^2_{H}$ is positive, which implies the non-existence of SU(2)
$\times $ U(1)
breaking minima. On the other hand  for $m_{3/2}\zzbar \le Q_0,
m^2_{H}< 0$, and so an SU(2) $\times$ U(1) breaking minimum is
developed with $m_{3/2}\zzbar$ near the transmutation scale $Q_0$.
These results are obtained by a minimization of the one-loop
effective potential in the $H_i$ and $m_{3/2}\zzbar$ field directions
[\ref{nsm}, \ref{fkpz}]. One finds
\begin{equation}
m_Z \simeq m_{3/2}\zzbar
\le O(Q_0)~.
\end{equation}
The no-scale SU(2) $\times$ U(1) radiative breaking provides the {\em
only
known mechanism} that explains the important hierarchy between the
electroweak scales $m_Z \sim m_{3/2} \le Q_0$ and the fundamental
gravitational scale $\mpl$.  This scale hierarchy is a consequence of
the very slow variation (logarithmic) of the renormalized
soft-breaking parameters:
\begin{equation}
Q_0 = \mpl \exp - \frac{O(1)}{\alpha_t} \simeq O(10^{-15})\mpl~.
\end{equation}
where $\alpha_t=h^2_t/4\pi$ is the top Yukawa coupling at $M_z$. The
$O(1)$ coefficient is determined in any specific no-scale model
and depends only on dimensionless parameters, gauge and Yukawa
coupling constants and on the soft- breaking terms mass ratios,
$m_{3/2}/M_{1/2}$, $\mu/M_{1/2}$, $A/M_{1/2}$, $B/M_{1/2}$.

In the framework of no-scale models, the SUSY spectrum at low
energies is more constrained than in a general $N=1$ supergravity
model.  It depends on at least one parameter less than the general
$N=1$ models, since a linear combination of $m_{3/2}$ and $M_{1/2}$,
is determined by the minimization of the effective
potential.  In practice, the no-scale models are even more
restrictive, because of the flatness requirement in the
$z^a$-direction of the scalar potential.  This requirement
drastically restricts the choice of interesting $N=1$ supergravity
models.  As we already stressed, the main assumption of no-scale
models is about the quantum gravitational corrections.  In order to
check this assumption, it is necessary to go even beyond the $N=1$
supergravity framework, and try to find theories in which the quantum
gravitational corrections make sense and behave like $N=1$ no-scale
supergravities for energies $E <\mpl$.

It is remarkable that such an extension exists in the framework of
$N=1$
four-dimensional superstrings [\ref{fds1}--\ref{fds6}].  In that
framework, the quantum gravitational corrections are under control,
and all interactions (gauge, Yukawa and
gravitational) are unified at the string scale $M_S \simeq mpl$.  At
the very
beginning of the string revolution, it was noticed that there existed
a connection between the superstring effective theories
[\ref{effcl}--\ref{ss2}] and the $N=1$ no-scale supergravity models.
Since then, many candidates were found and confirmed their no-scale
structure.  However, for technical reasons, which are related
to the complexity of string solutions, as well as to the ambiguities
related to the supersymmetry- breaking mechanism on strings (or in
string- effective theories), we do not have up to now precise
quantitative predictions;  we rather have qualitative predictions,
which are not, in fact, more restrictive than the no-scale effective
theories.

\section{ 4d-Superstrings and their Effective
Theories}

Superstring theory extents the validity of quantum field theory to
very short distances and defines consistently the quantization of all
interactions.
It is thus appropriate to try to investigate the behavior of string
dynamics at high energies and in regions of spacetime where the
gravitational field is strong.  There are several problems in gravity
where the classical, and even worse the semiclassical treatment have
perplexed physicists for decades. We are referring here to questions
concerning the behavior in regions of strong (or infinite) curvature
with both astrophysical (black holes) and cosmological (big-bang,
wormholes) interest. It is only appropriate to try to elucidate such
questions in the context of stringy gravity. There has been progress
towards this direction, and by now we have at least some ideas on how
different string gravity can be from general relativity in regions of
space- time with strong curvatures [\ref{strgrav}]. We need however
exact classical solutions of string theory in order to have more
quantitative control on phenomena that are characteristic of stringy
gravity.

On the other hand the special characteristics of superstrings do not
only reflect in the gravitational sector of the theory. There are
also important implications for particle physics, namely, concerning
the string low energy predictions at $M_Z$ and at the accessible by
the future, energy scale of ${\cal O}(1)$TeV. The first main property
of superstings is that they are ultraviolet finite theories (at least
perturbatively). The second important property is that  they unify
gravity with all other interactions. There are several ways to
construct four- dimensional superstrings with $N=1$ space time
supersymmetry. It is interesting however, that all N=1 superstring
constructions shows some {\em universality} properties and thus they
define a {\em special} class of $N=1$ supergravity theories. The
bosonic part of their effective $N=1$ supergravity action, restricted
up to two space-time derivatives, reads
$$
S^{\rm eff}_{\rm bos}=\int d^4 x{\sqrt {-g}} \big\{ \frac{1}{2}{\cal
R}-\frac{\nabla_{\mu} S \nabla^{\mu}{\bar S}}{(S+{\bar
S})}-\frac{\delta {\hat c}}{2}
$$
$$
-K_{I{\bar J}}(z^I )\nabla_{\mu} z^I \nabla^{\mu}{\bar z}^{\bar
J}-\frac{V(z^I)}{(S+{\bar S})}
$$
\begin{equation}
-\frac{k^a}{4}(S+{\bar S}) F^{a}_{\mu \nu}~F_{a}^{{\mu \nu}} +
\frac{ik^a}{4}(S-{\bar S})F^{a}_{{\mu \nu}}~{\tilde F}_{a}^{{\mu
\nu}}) \big\}
\label{Seff}
\end{equation}
where $S+{\bar S}$ is the dilaton field and $S-{\bar S}$ the
pseudoscalar axion (dual to the two-index antisymmetric tensor).
$F^{a}_{\mu\nu}$ are
the field strengths of the gauge bosons, and ${\tilde
F}^{a}_{\mu\nu}$ are
their duals. The dilaton potential which is proportional to the
central charge deficit $\delta {\hat c}$ is zero for the critical
strings and is different from zero for the non-critical string
solutions which are defined in curved space-time. For the $N=1$
superstring solutions $\delta {\hat c}=0$ and the scalar manifold
metric $K_{I{\bar J}}$  is given in terms of derivatives of a
K\"alher potential $K(z^I,{\bar z}^{\bar I})$, $K_{I{\bar J}}\equiv
\partial_I \partial_{\bar J}K(z^I,{\bar z}^{\bar I})$.
 For a given string solution, the gauge
group and the multiplet content are uniquely specified, and  so are
the K\"ahler and the gauge kinetic functions, which do indeed exhibit
the remarkable geometrical properties. Moreover, as an effect of the
string unification of all interactions, these
theories do not contain any explicit mass parameter besides the
string
mass scale $M_S$, in the sense that all couplings and masses of the
low-energy effective theory  are associated with the VEVs of some
moduli fields.
There is a vast literature concerning the effective supergravities
corresponding to four-dimensional superstring models with unbroken
$N=1$ local supersymmetry, both at the classical
[\ref{effcl}--\ref{ss2}]
and at the quantum [\ref{effqu}] level. The typical structure which
emerges is the following. The vector multiplets are fixed by the
four-dimensional gauge group characterizing the given class of string
solutions. As for the
chiral multiplets, there is always a universal `dilaton-axion'
multiplet, $S$, singlet under the gauge group, which at the classical
level entirely determines the gauge kinetic function,
\be
\label{gkf}
f_{ab} = k^a\delta_{ab} S \, .
\ee.
The above form of the string gauge kinetic function implies a
tree-level unification    relation at the string scale. This
unification does not include  only the gauge interactions but also
the  Yukawa  ones as well as the interactions among the scalars.
\be
\frac{1}{\alpha_i}=\frac{k^i}{\alpha_{str}}=2\pi k^i(S+{\bar
S})~~{\rm at}~~M_S
\ee
This unification of couplings happens  at  large energy scales
$E_t={\cal O}(M_{S})=5\times10^{17}$~GeV. At low scales due to the
quantum corrections,the above string unification relations become
\begin{equation}
\frac{1}{\alpha_i(\mu)}=\frac{k^i}{\alpha_{str}}+
b_0^i~log\frac{M^2_{S}}{\mu^2}+\Delta_{i}(T^a)
\label{stru} \end{equation}
The string unification looks very similar to the well-known
unification
condition in supersymmetric Grand Unified Theories (susy-GUTs) with a
unification scale  $M_U\sim M_S$ and $\Delta_{i}(T^a)=0$ in the
${\bar {DR}}$ renormalization scheme; in susy-GUTs the normalization
constants $k_i$ are fixed $only$ for the gauge couplings
($k_1=k_2=k_3=1$, $k_{em}=\frac{8}{3}$), but there are no relations
among gauge and Yukawa couplings at all. In string
effective theories, however, the normalization constants ($k_i$) are
known for both {\em gauge and Yukawa interactions}. Furthermore,
$\Delta_{i}(T^a)$ are calculable $finite$ quantities for any
particular string solution without any Ultra-Violet ambiguities since
strings are UV finite. Thus, the predictability of a given string
solution is extended for all low energy coupling constants
${\alpha_i(M_Z)}$ once the string- induced corrections
$\Delta_{i}(T^a)$ are determined. This determination  however,
requests string computations which we did not know, up to now, how to
perform in full generality. It turns out that $\Delta_{i}(T^a)$ are
non-trivial functions of the vacuum expectation values  of  some
gauge singlet fields the so-called moduli as well as standard  Higgs
fields [\ref{moduli}, \ref{dfkz}, \ref{ant}, \ref{infr}], (the moduli
fields are flat directions at the string classical level and they
remain flat in string perturbation theory, in the exact
supersymmetric limit). The $\Delta_{i}(T^a)$ are target space duality
invariant functions, which depend on the particular string solution.
Partial results for $\Delta_{i}$ exist [\ref{moduli}, \ref{dfkz},
\ref{ant}, \ref{infr}] in the exact
supersymmetric limit in many string solutions based on orbifold
[\ref{fds3}] and fermionic constructions [\ref{fds4}] and they are,
in principle,   well defined calculable quantities once we perform
our calculations  at the string
level  where all interactions including gravity are  consistently
defined. The full string corrections   to the coupling constant
unification, $\Delta_{i}(T^A)$, as well as the string corrections
associated to the soft supersymmetry-breaking parameters

\centerline{$m_0$, $m_{1/2}$, $A$, $B$ and $\mu$, at $M_U$,}

\noindent
are of main importance, since they fix  the strength of the gauge and
Yukawa interactions, the full spectrum of the supersymmetric
particles as well as the SM Higgs and the top-quark masses at the low
energy range $M_Z\leq E_{t}\leq {\cal O}(1)$ TeV.

 In addition to $S$, there are in general other singlet chiral
superfields, called `moduli', which do not appear in the
superpotential of the string effective theories and thus correspond
to classically flat directions of the scalar potential. They
parametrize the size and the shape of the internal compactification
manifold, and will be denoted here by the generic symbols $T$ and
$U$. Finally, there are other chiral superfields which are in general
charged under the gauge group, or at least have a potential induced
by some superpotential coupling: for the moment, we shall denote them
with the generic symbol $C$, understanding that in realistic models
this class of fields should
contain the matter and Higgs fields of the MSSM. ($z^I=[T,U,C]$).

The remarkable fact is that in the known four- dimensional string
models, in the limit where the $T$ and/or $U$ moduli are large with
respect to the string scale $M_S$, the K\"ahler manifold for the
chiral superfields obeys well- defined scaling properties with
respect to the real combinations of moduli fields $s \equiv
(S+\ov{S})$, $t_i \equiv (T_i+\ov{T}_i)$ and $u_i \equiv
(U_i+\ov{U}_i)$. These scaling properties are  due to the discrete
target- space duality symmetries [\ref{duality}]  of four-dimensional
superstrings, and the scaling weights are nothing but the modular
weights with respect to the moduli fields   which participate in the
supersymmetry-breaking mechanism: in the  limit of large moduli,
non-trivial topological effects on the  world sheet are exponentially
suppressed and can be neglected;  in this limit  the discrete duality
symmetries are promoted to accidental  scaling symmetries of the
kinetic terms in the effective  supergravity theory. More precisely,
the K\"ahler potential can be written as
\be
\label{kalstr}
K = - \log Y (s,t,u)
+ K^{(C)} (C,\ov{C};T,\ov{T};U,\ov{U}) \, .
\ee
The function $Y$ factorizes into three terms,
\be
\label{yfact}
Y = Y^{(S)} (s) \cdot Y^{(T)} (t) \cdot Y^{(U)} (u) \, ,
\ee
where
\be
Y^{(S)} = s \, ,
\ee
so that
\be
\label{scals}
s \partial_s Y = Y \, .
\ee
Another general feature involves the moduli $T_i$, corresponding to
harmonic $(1,1)$ forms, associated with deformations of the K\"ahler
class of the internal compactified space. Even if their number is
model-dependent, the
fact that three of them are related to the three complex coordinates
of the internal compactification manifold implies, in the limit of
large $T$ moduli,
\be
\label{scalt}
t_i \partial_{t_i} Y = 3 \, Y \, .
\ee
The moduli $U_i$ are associated with harmonic $(1,2)$ forms,
correspond to deformations of the complex structure of the internal
compactified space, and their existence, number and properties are
more model-dependent. In general, in the limit of large $U$ moduli
one can write a relation of the form
\be
\label{scalu}
u_i \partial_{u_i} Y = p_U \, Y \, ,
\ee
where $p_U=0,1,2,3$ depends on the superstring model under
consideration.
Finally, keeping only quadratic fluctuations of the $C$ fields
(sufficient to evaluate the K\"ahler metric and the mass
terms around $C = \ov{C} = 0$), one can in general write
$$
K^{(C)} = \sum_A K^A_{i_A \ov{j}_A} (t,u) C^{i_A}
\ov{C}^{\ov{j}_A}
$$
\be
{}~~~~~~+\label{kmatter}
{1 \over 2} \sum_{A,B} \left[  P_{i_A j_B}
(t,u) C^{i_A} C^{j_B} + {\rm h.c.} \right] + \ldots \, ,
\ee
with generic scaling properties of the form
\be
\label{scalct}
t_i \partial_{t_i} K^A_{i_A \ov{j}_A} =
\lambda_t^A K^A_{i_A \ov{j}_A}
\, ,
\ee
\be
\label{scalcu}
u_i \partial_{u_i} K^A_{i_A \ov{j}_A} =
\lambda_u^A K^A_{i_A \ov{j}_A}
\, ,
\ee
\be
\label{scalpt}
t_i \partial_{t_i} P_{i_A j_B} = {\lambda_t^A + \lambda_t^B \over 2}
P_{i_A j_B} \, ,
\ee
\be
\label{scalpu}
u_i \partial_{u_i} P_{i_A j_B} = {\lambda_u^A + \lambda_u^B \over 2}
P_{i_A j_B} \, .
\ee
These remarkable scaling properties for the K\"ahler metric follow
from the discrete target-space dualities, which are symmetries of the
full
K\"ahler function $\cg$. Under a generic duality transformation, of
the form
$$
z^\alpha \longrightarrow f(z^\alpha)\, ,
$$
the K\"ahler potential transform as
$$
K \longrightarrow K + \phi + \ov{\phi} \, ,
$$
where $\phi$ is an analytic function of the moduli fields $z^\alpha$,
and in particular it must be that
\be
Y \longrightarrow Y e^{\phi + \ov{\phi}} \, .
\ee
Also, it is not restrictive for our purposes to consider the case in
which
the fields $C_A$ transform with a specific modular weight
$\lambda_A$,
\be
C_A \longrightarrow e^{- \lambda_A \phi} C_A \, .
\ee
The fact that target-space duality is a symmetry implies then a
definite transformation property for the superpotential,
$w \longrightarrow e^{- \phi} w $ which in turn puts very strong
restrictions on the superpotential couplings, for example the cubic
Yukawa couplings of the form $~h_{A B D} C^{A} C^{B} C^{D}$. If $h_{A
B D}$ is such that, in the large moduli limit, it goes to a
non-vanishing constant (or, more generally, to a modular form of
weight zero), then it must be
\be
\lambda_A + \lambda_B + \lambda_D = 1 \, .
\ee
For example, in $Z_2 \times Z_2$ orbifolds the $h_{A B D}$ are
constants, whereas in Calabi-Yau manifolds they are modular forms of
weight
zero, which approach a constant in the large volume limit for the
associated
moduli.

In the case of unbroken supersymmetry, and in the large moduli limit,
the classical superpotential $w$ is independent of the $(S,T,U)$
moduli fields, and at least quadratic in the $C$ fields. From the
previous scaling properties, it also follows that around $C=0$ one
can write
\be
K^s K_s = 1 \, ,
\;\;\;\;
K^{t_i} K_{t_i} = 3 \, ,
\;\;\;\;
K^{u_i} K_{u_i} = p_U \, .
\ee
Armed with this result, we are ready to discuss spontaneous
supersymmetry breaking in the superstring effective supergravities.
As already explained, to have broken supersymmetry and vanishing
vacuum
energy one needs $w \ne 0$ and $\cg^I \cg_I = 3$ at the minima of the
tree-level potential. If one takes the effective supergravities
derived from the four- dimensional superstring model with unbroken
supersymmetry, one consistently obtains a semi- positive definite
scalar potential, admitting $C=0$ minima with unbroken supersymmetry
and vanishing
vacuum energy, and flat directions along the $S$, $T$ and $U$ moduli
fields. To obtain supersymmetry breaking minima with unbroken gauge
symmetries,  one has to introduce a superpotential modification which
generates
minima with $C=0$, $w \ne 0$, $\cg^I \cg_I = 3$ when the summation
index $I$ runs over the ($S,T,U$) moduli, $\cg^I \cg_I = 0$ when $I$
runs over the $C$ fields. This means, however, that the
superpotential modification must depend on at least some of the
($S,T,U$) moduli, since otherwise we would get, when summing over the
moduli indices, $\cg^I \cg_I = 4 + p_U$, which would make the scalar
potential strictly positive definite and thus not allow for the
desired minima.  As for the origin of possible superpotential
modifications, we must refer to the two types of mechanisms for
supersymmetry breaking considered so far in the framework of four-
dimensional string models. The first one corresponds to exact tree-
level string solutions [\ref{ssbr1}, \ref{ssbr2}], in which
supersymmetry is broken via orbifold compactification. The second one
is based on the assumption
that supersymmetry breaking is induced by non-perturbative phenomena
[\ref{gcond}], such as gaugino condensation or something else, at the
level of the string effective field theory. These will be the two
possibilities considered in the following.

In the case of non-perturbative supersymmetry breaking [\ref{gcond}],
in the absence of a second-quantized string formalism one can assume
that, at the level of the effective supergravity, the super- Higgs
mechanism is induced by a superpotential modification which preserves
target-space duality [\ref{flst}].
The relevant transformations are those acting non-trivially on the
moduli fields $z^\alpha$ associated with supersymmetry breaking. If,
for
example, the modified superpotential has the form
\be
w = w_{SUSY} + A(z^\alpha) + B_{A B}(z^\alpha) C^{A} C^{B}
+ \ldots \, ,
\ee
target-space duality requires then the following transformation
properties:
\be
A(z^\alpha) \longrightarrow A(z^\alpha) e^{-\phi} \, ,
{}~~~~~~~~~~~~
B_{A B}(z^\alpha) \longrightarrow B_{A B}(z^\alpha) e^{-(1 -
\lambda_A - \lambda_B)\phi} \, .
\ee
Unfortunately, the form of the function $A(z^\alpha)$ cannot be
uniquely fixed by the requirement that it is a modular form of weight
$(-1)$.
However, another important constraint comes from the physical
requirement
that the potential must break supersymmetry and generate a vacuum
energy at
most ${\cal O}(m_{3/2}^4)$ in the large moduli limit. This implies
that
$A(z^\alpha) \longrightarrow {\rm constant} \ne 0$ for $z^\alpha
\rightarrow \infty$. This is not the case for the models of
supersymmetry
breaking with minima of the effective potential at small values of
$T$,
which make use of the Dedekind function $\eta(T)$ in the
superpotential
modification [\ref{lust}]: either they do not break supersymmetry or
they
do so with a large cosmological constant, in  contradiction with the
assumption of a constant flat background.
In the case of the function  $B_{A B}(z^\alpha)$,
it is sufficient to assume that, in the large moduli limit, $B_{A
B}(z^\alpha) \longrightarrow {\rm constant}$. For the moduli fields
that are not involved in the breaking of supersymmetry, these
asymptotic conditions are not necessary and can be relaxed.
The requirement that $A(z^\alpha) \longrightarrow {\rm constant} \ne
0$ for $z^\alpha \rightarrow \infty$ defines an approximate no-scale
model, with minima of the effective potential corresponding to
field configurations that are far away from possible $z^\alpha
\simeq {\cal O}(1)$ self-dual minima with unbroken supersymmetry
($\cg_\alpha =0$) and negative vacuum energy ${\cal O}(M_P^4)$.
Between these two classes of  extrema, there may exist other extrema
of the effective potential with $\cg_\alpha \ne 0$, but those are
generically unstable and/or have non-vanishing vacuum energy
[\ref{lust}]. As for the VEVs of the moduli fields that do not
contribute to
supersymmetry breaking (those with $\cg^i \cg_i = 0$), they are
generically fixed to some extended symmetry points (e.g. the
self-dual points).

In the string models with tree-level supersymmetry-breaking
[\ref{ssbr1}, \ref{ssbr2}], the
superpotential modifications in the large-moduli limit are fully
under control, since in that case the explicit form of the one-loop
string partition function is known, and one can derive the low-energy
effective theory without making any assumption.  In this class of
models, the large-moduli limit is a necessity, since for small values
of the moduli (close to their self-dual points) there exist
Hagedorn-type instabilities, induced by some winding modes that
become tachyonic in flat space-time [\ref{ssbr2}]. At the self-dual
point there is a new stable minimum with unbroken supersymmetry and
negative cosmological constant, as expected. We should stress here
that the prescription
for a consistent effective field theory in the region of small moduli
requires the addition of extra degrees freedom, corresponding to the
winding
modes which can become massless or tachyonic for some values of the
$T$ and/or $U$ moduli close to the self-dual points [\ref{ssbr2}]. In
the large-moduli
limit, however, we can disregard the effects of these extra states
and not include them in the effective field theory. In this limit,
the superpotential modification associated with supersymmetry
breaking seems to violate target-space duality. On the other hand,
$w_{SUSY}$
and the K\"ahler potential maintain the same expressions as in the
case of
exact supersymmetry, with the desired scaling properties that can
produce a
supergravity model with $Q\zzbar=0$ [\ref{fkpz}].
 The main features of the effective theories with tree- level
supersymmetry breaking are the following:
\begin{enumerate}
\item
The K\"ahler potential and the gauge kinetic function of the
effective theory are the same as those obtained in the limit of
unbroken supersymmetry, so that, up to analytic field redefinitions,
supersymmetry breaking is indeed
induced only by a superpotential modification.
\item
For $C=0$, the K\"ahler manifold for the $T$ and $U$ moduli can be
decomposed into the product of two factor manifolds. The first one,
described by a K\"ahler potential $K'$, involves one $T$ and
one $U$ field, to be called here $T'$ and $U'$
\be
K' = - \log [(T' + \ov{T}')(U' + \ov{U}')] \, ,
\;\;\;\;\;
(C=0) \, ,
\ee
and the second one, described by a K\"ahler potential $K''$, involves
all the remaining $T$ and $U$ moduli.
\item
The superpotential modification associated with supersymmetry
breaking does not involve the fields $S$, $T'$ and $U'$, so that
$\cg_S=K_S$, $\cg_{T'}=K_{T'}$, $\cg_{U'}=K_{U'}$. The condition
$\cg^{\alpha} \cg_{\alpha} = 3$, which must be satisfied at the
minima, is identically saturated by the fact that for $C=0$ it is
$\cg^{S} \cg_{S} = \cg^{T'} \cg_{T'} = \cg^{U'} \cg_{U'} = 1$. The
goldstino direction is then along some linear combination of the
$(S,T',U')$ fields.
\item
The superpotential modification associated with supersymmetry
breaking involves the fields appearing in $K''$, so that,
restricting the sum over $I$ to these fields, the condition
$\cg^I \cg_I = 0$ can be satisfied at all minima.
\end{enumerate}

We can now discuss the other proposed mechanism for spontaneous
supersymmetry breaking in string- derived supergravity models, i.e.
the possibility of non- perturbative phenomena, which at the level of
the effective supergravity theory can again be described by a
modification of the superpotential [\ref{gcond}].
In order to obtain a consistent model, with broken supersymmetry and
classically vanishing vacuum energy, the superpotential modification
must be such that $\cg^I \cg_I = 3$ around $C=0$.
The superpotential
modification must then contain some dependence on the $(S,T,U)$
moduli
fields in order to avoid a strictly positive potential.
For example, the simplest choice $w= k + O(C^2)$, with $w$
independent of the $(S,T,U)$ moduli, would give $\cg^I \cg_I =
K^{\hat{I}} K_{\hat{I}} = 4 + p_U$, where the index $\hat{I}$ runs
over the $(S,T,U)$ moduli, and therefore a potential around $C=0$ of
the form $V = (1+p_U) |k|^2 e^K=(1+p_U)|k|^2 / Y$.
Since the quantity at the numerator is field-independent and strictly
positive-definite, there is no stationary point for the potential,
with the exception of the boundaries of moduli space, $Y \rightarrow
\infty$, for example the decompactification limit $Y^{(T)} Y^{(U)}
\rightarrow \infty$ or the zero-coupling limit $Y^{(S)} \rightarrow
\infty$. It is then clear that, to avoid this problem, the
superpotential must depend on some of the moduli, in order to fix
some of the VEVs associated with the moduli directions.

The simplest superpotential modification follows from the conjecture
of gaugino condensation [\ref{gcond}], and includes a non-trivial
dependence on the
$S$ modulus. Such an assumption is made plausible by the fact that
the
gauge coupling constant of the theory is determined by the VEV of the
$S$ field. An $S$-dependent superpotential modification can allow for
minima with $\cg_S = 0$, and fix the VEV of the $S$ modulus at the
minima. Irrespectively of the details of the $S$ dependence of the
superpotential, as long as there is a field configuration of $S$
such that $\cg_S=0$, this is sufficient to create a well-behaved
positive-semi-definite potential in the absence of $U$-type moduli
($p_U=0$). When such moduli are present, one must make the further
assumption that the superpotential contains also a non-trivial
$U$-dependence, so that minima with $\cg_U=0$ can be allowed:
otherwise, the scalar potential would still remain strictly
positive-definite. Notice that the stabilization of the VEVs of the
$U$-type
moduli can be performed either at the string level, by moving to the
points of extended symmetry associated with the $U$ moduli, or at the
level of the effective theory, by extending the assumption made for
the $S$ field.

A superpotential modification with non-trivial $S$ and $U$ dependence
has the name of `$T$-breaking', since in that case the condition of
vanishing vacuum
energy, $\cg^I \cg_I = 3$, is saturated by the $T$ fields only. For
the models in which $p_U=3$, we may alternatively assume a
superpotential modification, which depends only on $S$ and $T$, so
that $\cg_S = \cg_{T_i} = 0$ at the minima, and the condition $\cg^I
\cg_I = 3$ is entirely saturated by the $U$ moduli: we shall call
this scenario `$U$-breaking'. In the cases in which the K\"ahler
manifolds
for the $T$ and $U$ moduli are factorized, one may also consider
intermediate scenarios of $S/T/U$-breaking, in which the
superpotential modification is such that, at the minima of the
potential with $C=0$, it is identically $\cg^S \cg_S + \cg^{T_i}
\cg_{T_i} + \cg^{U_i} \cg_{U_i} = 3$, with non-vanishing
contributions from more than one sector. In this language the string
tree level breaking [\ref{ssbr1}, \ref{ssbr2}] is an $S/T'/U'$-
breaking. I should stress here that in all these scenarios the
resulting value of $Q$, the coefficient appearing in $\str \cm^2$,
does not depend on the details of the superpotential
modification [\ref{fkpz}], but only on the scaling weights
$\lambda^I$ of the different fields $z^I$, with respect to the moduli
$z^\alpha$ for which $\cg^{\alpha}
\cg_{\alpha} \ne 0$ (not summed) at the minima.
\be
z^{\alpha} \partial _{\alpha}[K_{I,{\bar J}}]^C=\lambda^C
\ee
What is interesting is that under the above assumptions for the
supersymmetry breaking, the coefficient  $Q$ and the soft breaking
parameters are given {\em uniquely} in terms of the scaling weights
$\lambda^I$ and in terms of the gravitino mass scale $m_{3/2}$; there
is no  any further dependence coming from the superpotential
[\ref{fkpz}].\\
$\bullet$ Quadratic divergences:
$$
Q=\sum_A (1+\lambda^A)+ d^a{\lambda^a}-1
$$
$\bullet$ Scalar mass terms:
$$
(m_0^2)_A=(1+\lambda^A)m_{3/2}^2
$$
$\bullet$ Gaugino mass terms:
$$
(M_{1/2})_a=-\lambda^a m_{3/2}
$$
$\bullet$ Cubic Scalar Couplings:
$$
A_{A,B,C}=(3+\lambda^A+\lambda^B+\lambda^C) m_{3/2}
$$
$\bullet$ $\mu_{A,B}~H_A~H_B$ terms:
$$
\mu_{A,B}=\frac{1}{2}(2+\lambda^A+\lambda^B) m_{3/2}
$$
$\bullet$ Quadratic Scalar Couplings:
$$
B_{A,B}=\frac{1}{2}(4+\lambda^A+\lambda^B) m_{3/2})
$$
In the above equations $\lambda^a$ is the scaling weight of the
gauge
kinetic function and $d^a$ is the dimension of the gauge group.

\section{Superstrings Corrections to the Unification Relations}

The main obstruction in determining the exact form of the string
radiative corrections $\Delta_{i}(T^a)$ is strongly related to the
infrared divergences of the $\langle [F^a_{\mu\nu}]^2\rangle $
two-point correlation function in superstring theory. In field
theory, we can avoid this problem using off-shell calculations. In
first
quantized string theory we cannot do so since we do not know in
general how to
go off-shell. Even in field theory there are problems in defining an
infrared
regulator for chiral fermions especially in the presence of space-
time supersymmetry.

In [\ref{cw}] it was suggested to use a specific space- time with
negative curvature in order to achieve consistent regularization in
the infrared. The proposed  curved space however is not useful for
string applications since it does not correspond to an exact
super-string solution.

Recently, exact superstring solutions have been constructed  using
special four-dimensional spaces as superconformal building blocks
with
${\hat c}=4$ and $N=4$ superconformal symmetry [\ref{n4kounnas},
\ref{worm}, \ref{infr}]. The full spectrum of string excitations for
the superstring solutions based on those
four-dimensional subspaces, can be derived using the techniques
developed in [\ref{worm}]. The main characteristic property of these
solutions is the existence of a mass gap $\mu$, which is proportional
to the curvature of the non-trivial  four-dimensional space- time.
Comparing the spectrum in  a flat background with that in curved
space we observe a shifting of all massless states by an amount
proportional to the space- time curvature, $\Delta
m^2=Q^2/4=\mu^2/2$, where $Q$ is the Liouville background charge and
$\mu$ is the IR cutoff. What is also interesting is that the shifted
spectrum in curved space is equal for bosons and fermions due to the
existence of a new
space-time supersymmetry defined in curved space- time
[\ref{n4kounnas}, \ref{worm}, \ref{infr}]. Therefore, in some curved
space- time solutions the induced infrared regularization is
consistent with supersymmetry and can be used either in field theory
or string theory.

In order to calculate the
renormalization of the effective couplings we need to turn on
backgrounds for gauge and gravitational fields.
Thus our aim is to define the  deformation of the two-dimensional
superconformal theory  which corresponds to a non-zero field strength
$F^{a}_{\mu\nu}$ and $R_{\mu\nu\rho\sigma}$
background
and find the integrated  one-loop
partition function $ Z(\mu,F,{\cal R})$,  where $F$ is by the
magnitude
of the field strength,
$F^2 \equiv \langle F^{a}_{\mu\nu}F_{a}^{\mu\nu}\rangle$ and $\cal R$
is
that of the curvature,  $\langle
R_{\mu\nu\rho\sigma}R^{\mu\nu\rho\sigma}\rangle={\cal R}^2$.
\begin{equation}
Z[\mu,F_i,{\cal R}]=\frac{1}{V(\mu)} \int_{\cal F}
\frac{ d\tau d{\bar\tau} }{ ({\rm Im}\tau)^2 }
Z[\mu,F_i,{\cal R};\tau,{\bar\tau}]
\label{intpart}
\end{equation}
The index $i$ labels different simple or $U(1)$ factors of the gauge
group and $V(\mu)=(k+2)^{3/2}/8\pi$ is the volume of the three
dimensional
sphere. Expanding the partition function in a power series in $F,
{\cal R}$
\be
Z[\mu,F_i,{\cal R}]=\sum_{m,n=0}^{\infty}F^m{\cal R}^n
\ee
we can extract the integrated correlators $\langle F^m {\cal
R}^n\rangle
=Z_{m,n}$. The one loop correction to the gauge coupling constants is
given by [\ref{infr}]
\be
{4\pi\over \alpha_{A}^2(\mu)}={{4\pi k^a}\over
\alpha_{a}^2(M_{S})}+Z^{a}_{2,0}(\mu)
\label{cc}
\ee
 In flat space, a small non-zero  $F_{\mu\nu}^a$ background gives
rise
to an infinitesimal deformation  of the 2-d $\sigma$-model action
given by,
\begin{equation}
\Delta S^{2d}(F^{(4)})=\int dzd{\bar z}\;F_{\mu\nu}^a[x^{\mu}
\partial_z x^{\nu}+\psi^{\mu}\psi^{\nu}]{\bar J}_a
\label{fdef}
\end{equation}
Observe that for $F^a_{\mu\nu}$ constant (constant magnetic field),
the left moving operator $[x^{\mu} \partial_z
x^{\nu}+\psi^{\mu}\psi^{\nu}]$ is not a well-defined $(1,0)$ operator
on the world sheet. Even though  the right moving Kac-Moody current
${\bar J}_a$ is a well-defined $(0,1)$ operator, the total
deformation
is not integrable in flat space. Indeed, the 2-d $\sigma$-model
$\beta$-functions are not satisfied in the presence of a constant
magnetic field. This follows from the fact that there is a
$non$-$trivial$ $back$-$reaction$ on the gravitational background due
the non-zero
magnetic field.

The important property of a non-trivial space- time background in
which the 4d
flat background $R^4$ is replaced by $ W^{(4)}_k =S^3_k\times R$, a
three dimensional sphere with curvatue $1/(k+2)$ plus a
non compact coordinate with a background $Q=\sqrt{2/(k+2)}$ , is
that we can solve exactly for the gravitational
back-reaction. First observe that the deformation that
corresponds to a constant magnetic field
$B_i^a=\epsilon_{oijk}F_a^{ik}$ is a well-defined
(1,1) integrable deformation,  preserving the world-sheet
supersymmetry:
\begin{equation}
\Delta S^{2d}(W^{(4)}_k)=\int dzd{\bar
z}\;B^a_i[I^i+\frac{1}{2}\epsilon^{ijk}\psi_{j}\psi_{k}]{\bar J}_a
\label{fdef1}
\end{equation}
where $I^i$ is anyone of the $SU(2)_{k}\sim S^3_k$ currents.
The deformed partition function $Z[\mu,F_i,{\cal R}]$ is IR finite
due to the infrared regulator $\mu$ induced by the weakly curved
space- time. In the string framework it is also UV finite, where the
$M_S$ acts as UV cutoff.

In general the exact determination of the IR regularized partition
function in terms of the expectation values of the moduli fields
$T^I$, the Higgs fields $H^I$ and the supersymmetry auxiliary fields
 $\cal G^I$,
$Z[\mu,F_i,{\cal R},T^I,H^I,{\cal G^I}]$, defines at
one loop all string quantum corrections for the gauge, Yukawa and
gravitational couplings as well as the corrections to the soft
supersymmetry  breaking parameters.

\section{ Conclusions}

 In realistic models of
spontaneously broken supergravity, the desired hierarchy $m_Z,m_{3/2}
\ll \mpl$ can be stable, and eventually find a natural dynamical
explanation, when quantum loop corrections to the effective
potential do not contain terms quadratic in the cut-off scale $\mpl$.
 Requiring broken supersymmetry
with vacuum energy at most ${\cal O}( m_{3/2}^4)$ at one loop level,
 defines a highly non-trivial constraint on the K\"ahler potential
$K$ and the gauge kinetic function $f_{ab}$, including both the
observable and the
hidden sectors of the theory, as well as on the mechanism for
spontaneous supersymmetry breaking. In the presence of some
approximate scaling properties of the gauge and K\"ahler metrics the
contributions to the coefficient of the quadratic divergences $Q$ and
to the soft supersymmetry
breaking parameters depend
only on the scaling weights of the fields $\lambda_i$, and not on the
VEVs of the sliding singlet fields in the
hidden sector.
The expressions for $Q$ and for the mass parameters of the MSSM in
terms of $\lambda^i$
find a deeper justification in the effective theories of four-
dimensional
superstrings, where supersymmetry breaking is described either at the
string tree-level or, by assuming some non-perturbative phenomena,
in the effective field theory. In these theories, the full
particle content and
the approximate scaling weights are completely fixed. The origin of
the
approximate scaling properties of the superstring effective theories
is due to target-space modular invariance, and the scaling weights
$\lambda^i$ are nothing but the target-space duality weights with
respect to the moduli fields, which participate in the
supersymmetry- breaking mechanism.
Indeed, in the limit of large moduli the discrete
target-space duality symmetries are promoted to some accidental
scaling symmetries of the gauge and matter kinetic terms in the
effective supergravity theory.
Thus we have identified in $Q=0$ another criterion for a
consistent choice of the supersymmetry breaking directions $\cg^I \ne
0$ in a given 4d- superstring model. Once the breaking direction in
the auxiliary field space is identified $\cg^I \ne 0$, we can
construct the IR regularized partition function at one loop in the
presence of  non zero magnetic field $F^a$, non zero curvature $\cal
R$ and in terms of the expectation values of the moduli fields $T^I$
and the Higgs fields $H^I$
$$
Z(\mu, F^a, {\cal R}, T^I, H^I, \cg^I ).
$$
The knowledge of the above partition function defines at one loop all
string quantum corrections for the gauge, Yukawa and gravitational
couplings as well as the corrections to the soft supersymmetry
breaking parameters.

\vskip 1cm
{\bf Acknowledgements}

We would like to thank  the Organizing Committees of the SUSY-95
and Four Seas Conferences for giving me the opportunity to
present our results. I would also like to thank S. Ferrara, E.
Kiritsis
and F. Zwirner who collaborated in part of the work presented here.
This work was  supported in part by EEC contracts
SC1$^*$-0394C and SC1$^*$-CT92-0789.

\end{document}